\documentclass[preprint,11pt]{revtex4}
\usepackage{graphicx}
\usepackage{epstopdf}
\usepackage{amsmath}
\usepackage{amsfonts}
\usepackage{amssymb}
\usepackage{times}
\def\fo{\hbox{{1}\kern-.25em\hbox{l}}}

\newcommand{\newc}{\newcommand}
\newc{\lcal}{\int {\cal L}dt}
\newc{\LSP}{{\chi^0_1}}
\newc{\stauR}{{\tilde \tau_R}}
\newc{\stau}{{\tilde \tau_1}}
\newc{\mstop}{m_{\tilde{t}}}
\newc{\mHpm}{m_{H^\pm}}
\newc{\gsim}{\lower.7ex\hbox{$\;\stackrel{\textstyle>}{\sim}\;$}}
\newc{\lsim}{\lower.7ex\hbox{$\;\stackrel{\textstyle<}{\sim}\;$}}
\newc{\ie}{{\it i.e.}}
\newc{\etal}{{\it et al.}}
\newc{\eg}{{\it e.g.}}
\newc{\kev}{\hbox{\rm\,keV}}
\newc{\mev}{\hbox{\rm\,MeV}}
\newc{\gev}{\hbox{\rm\,GeV}}
\newc{\tev}{\hbox{\rm\,TeV}}
\newc{\xpb}{\hbox{\rm\, pb}}
\newc{\xfb}{\hbox{\rm\, fb}}

\newc{\mtop}{m_t}
\newc{\mbot}{m_b}
\newc{\mz}{m_Z}
\newc{\mw}{M_W}
\newc{\alphasmz}{\alpha_s(m_Z^2)}
\newc{\swsq}{\sin^2\theta_W}
\newc{\tw}{\tan\theta_W}
\newc{\cw}{\cos\theta_W}
\newc{\sw}{\sin\theta_W}
\newc{\BR}{\hbox{\rm BR}}
\newc{\zbb}{Z\to b\bar}
\newc{\Gb}{\Gamma (Z\to b\bar b)}
\newc{\Gh}{\Gamma (Z\to \hbox{\rm hadrons})}
\newc{\rbsm}{R_b^\hbox{\rm sm}}
\newc{\rbsusy}{R_b^\hbox{\rm susy}}
\newc{\drb}{\delta R_b}
\newc{\sgn}{\mbox{sgn}}
\newc{\tbeta}{\tan\beta}
\newc{\uL}{{\tilde u_L}}
\newc{\uR}{{\tilde u_R}}
\newc{\cL}{{\tilde c_L}}
\newc{\cR}{{\tilde c_R}}
\newc{\tL}{{\tilde t_L}}
\newc{\tR}{{\tilde t_R}}
\newc{\dL}{{\tilde d_L}}
\newc{\dR}{{\tilde d_R}}
\newc{\sL}{{\tilde s_L}}
\newc{\sR}{{\tilde s_R}}
\newc{\bL}{{\tilde b_L}}
\newc{\bR}{{\tilde b_R}}
\newc{\eL}{{\tilde e_L}}
\newc{\eR}{{\tilde e_R}}
\newc{\mhp}{m_{H^\pm}}
\newc{\mhalf}{m_{1/2}}
\newc{\emt}{{e/\mu /\tau}}
\newc{\lR}{\tilde{l}_R}
\newc{\lL}{\tilde{l}_L}
\newc{\nL}{\tilde{\nu}_L}
\newc{\na}{\chi^0_1}
\newc{\nb}{\chi^0_2}
\newc{\nc}{\chi^0_3}
\newc{\nd}{\chi^0_4}
\newc{\ca}{\chi^{\pm}_1}
\newc{\cb}{\chi^{\pm}_2}
\newc{\camp}{\chi^\mp_1}
\newc{\cbmp}{\chi^\mp_1}
\newc{\capos}{\chi^{+}_1}
\newc{\caneg}{\chi^{-}_1}
\newc{\phit}{\phi_t}
\newc{\phib}{\phi_b}
\newc{\phiew}{\phi_{ew}}
\newc{\htz}{h^0_t}
\newc{\hbz}{h^0_b}
\newc{\hewz}{h^0_{ew}}
\newc{\hsmz}{h^0_{sm}}
\newc{\huz}{h^0_u}
\newc{\hsusyz}{h^0_{susy}}
\def\mp{M_P}

\def\beq{\begin{equation}}
\def\eeq{\end{equation}}
\def\bea{\begin{eqnarray}}
\def\eea{\end{eqnarray}}

\begin{document}
\begin{center}
\vspace{3cm}
\end{center}
\title{\Large \bf Higgs Boson Masses in the MSSM with General Soft Breaking}
\author{Asl{\i} Sabanc{\i}$^{a}$,  Alper Hayreter$^{a}$, Levent Solmaz$^{b}$}
\affiliation{$^a$ Department of Physics, Izmir Institute of
Technology, IZTECH, Turkey, TR35430} \affiliation{$^b$ Department
of Physics, Bal{\i}kesir University, Bal{\i}kesir, Turkey,
TR10100}

\date{\today}
\begin{abstract}
\medskip
The operators that break supersymmetry can be
holomorphic or non-holomorphic in structure. The latter do not
pose any problem for gauge hierarchy and are soft provided that
the particle spectrum does not contain any gauge singlets. In
minimal supersymmetric model (MSSM) we discuss the impact of
non-holomorphic soft-breaking terms on the Higgs sector. We find
that non-holomorphic operators can cause significant changes as
are best exhibited by the correlation between the masses of the
charginos and Higgs bosons.

\end{abstract}

\maketitle

\section{Introduction}
In general, breakdown of global supersymmetry is parameterized by
a set of operators with dimensionality less than four. Each
operator thus comes with an associated mass scale, which must fall
in the TeV domain if supersymmetry is the correct description of
Nature beyond Fermi energies. The operators that break the
supersymmetry must be soft $i.e.$ quadratic divergences must not
be regenerated. The mass terms for scalars as well as their
trilinear couplings are soft operators \cite{lisa}. However, the
most general list of supersymmetry breaking operators involve
novel structures beyond these. Indeed, trilinear couplings, for
example, can have both $A \phi \phi \phi$ type holomorphic
structure as well as $A^{\prime} \phi^{\star} \phi \phi$ type
non-holomorphic structure. There is nothing wrong in considering
the non-holomorphic structures since they are perfectly soft if
there are no gauge singlets in the theory \cite{girardello}. In
this sense, MSSM provides a perfect arena for analyzing
consequences of the non-holomorphic soft-breaking terms.

In this work we study the impact of non-holomorphic soft-breaking
terms on the Higgs sector of the MSSM. We will analyze Higgs
sector in conjunction with the chargino sector so as to single out
the effects of non-holomorphic trilinear couplings from the $\mu$
parameter. Since these sectors are two of the prime concerns of
experiments at the LHC, we expect that our results will be
testable in near future.

The holomorphic soft terms in the MSSM involve
\begin{eqnarray}
\label{soft} -{\cal{L}}_{soft} \ni \widetilde{Q}\cdot {H}_u { {h_t
{A}_t}} \widetilde{U} + \widetilde{Q}\cdot {H}_d { {h_b{ A}_b}}
\widetilde{D} + \widetilde{L}\cdot {H}_d { {h_{\tau}{A}_{\tau}}}
\widetilde{E} + \mbox{h.c.}
\end{eqnarray}
in addition to the mass terms of scalars and gauginos. In writing
this equation, we have included only the third generation of
sfermions with trilinear couplings $A_{t, b, \tau}$.

As mentioned before, as has been shown explicitly in
\cite{girardello,poppitz}, in supersymmetric theories which do not
have pure gauge singlets in their particle spectrum, the holomorphic
supersymmetry breaking terms do not necessarily represent the most
general set of soft-breaking operators. Namely, the non-holomorphic
operators can also be regarded as soft symmetry breaking operators
as long as the theory does not contain any gauge singlet chiral
superfields. Being free of any gauge singlets, the soft breaking
sector of the MSSM can be enlarged to cover
\begin{eqnarray}
\label{softp} -{\cal{L}}_{soft}^{\prime} = \widetilde{Q}\cdot
{H}_d^{c} { {h_t {A}^{\prime}_t}} \widetilde{U} +
\widetilde{Q}\cdot {H}_u^{c} { {h_b{A}^{\prime}_b}} \widetilde{D}
 + \widetilde{L}\cdot {H}_u^{c} { {h_{\tau}{A}^{\prime}_{\tau}}} \widetilde{E} + \mbox{h.c.}
\end{eqnarray}
in addition to (\ref{soft}). Here ${ { A}^{\prime}_{t,b,\tau}}$
are non-holomorphic trilinear couplings which do not need to bear
any relationship to the holomorphic ones in ${ { A}_{u,d,e}}$ in
(\ref{soft}).

The trilinear couplings in (\ref{soft}) and (\ref{softp}) break
supersymmetry in a soft fashion, that is, in a way without
regenerating quadratic divergences, and hence, both set of operators
must be included in phenomenology of low-energy supersymmetry
\cite{cakir,lisa}.

Our concern in this letter is the Higgs sector. At low $\tan\beta$
($\tan\beta \leq 30$), radiative effects in the Higgs sector drive
mainly from the top (s)quarks since other fermions are too light to
have significant Yukawa interactions. Thus, radiative corrections to
Higgs sector \cite{higgs,higgs2,higgs3} may be computed by
considering an effective potential approach \cite{higgs2,higgs3}
with the top quark and scalar top quark loops. Their field-dependent
mass-squareds are given by
\begin{eqnarray}
\label{topmass} m_t^2(H) = h_t^2 \mid H_u^0\mid^2\,,
\end{eqnarray}
for top quarks,
\begin{eqnarray}
\label{stopmass} M_{\widetilde{t}}^{2}(H) = \left(\begin{array}{c c}
m_{\widetilde{t}_L}^2 + m_t^2 - \frac{1}{4} \left(g_2^2 -
\frac{1}{3} g_Y^2\right) \left(\mid H_u^0\mid^2 - \mid
H_d^0\mid^2\right) & h_t A_t^{\star}
H_{u}^{0\,\star} - h_t (\mu + { A}_t^{\prime\ \star}) H_d^{0}\\
h_t A_t H_{u}^{0} - h_t (\mu^{\star} + { A}_t^{\prime} ) H_d^{0\,
\star} & m_{\widetilde{t}_R}^2 + m_t^2 - \frac{1}{3} g_Y^2
\left(\mid H_u^0\mid^2 - \mid H_d^0\mid^2\right)\end{array}\right)
\end{eqnarray}
for top squarks. This very expression for stop masses clearly
shows that entire effect of including the non-holomorphic triliear
coupling $A_t^{\prime}$ is to shift the $\mu$ parameter as
\begin{eqnarray}
\label{shift} \mu \rightarrow \mu + A^{\prime}_t
\end{eqnarray}
which implies that all the effects of scalar top quarks on the
Higgs sector, as described in detail in \cite{higgs,higgs2,higgs3}
for holomorphic soft terms, remain intact except that the $\mu$
parameter is not as it is in the superpotential yet it
is the shifted one in (\ref{shift}).

Consequently, effects of the non-holomorphic trilinear coupling
$A^{\prime}$ parameter can be disentangled from those of the $\mu$
parameter if $\mu$ is known from an independent source. Clearly,
an independent knowledge of $\mu$ can be obtained from neutralino
or chargino sectors either via direct searches or via indirect
bounds from certain observables. A readily recalled observable is
$b \rightarrow s \gamma$ decay \cite{bsgam1,bsgam2,bsgam3,bsgam4}.
In addition one can consider bounds from EDMs or muon $g-2$ and
suchlike but for purposes of obtaining a simple yet direct
constraint on $\mu$--$A_{t}^{\prime}$ relationship $b\rightarrow s
\gamma$ decay suffices.

We first detail radiative corrections to Higgs boson masses in
light of (\ref{topmass},\ref{stopmass}) and then revisit the
$b\rightarrow s \gamma$ decay in order to show how $A_t^{\prime}$
can be extracted from the bulk of the MSSM parameters. As mentioned
before, the Higgs boson masses depend on $\mu + A_t^{\prime}$ not
$A_t^{\prime}$ in isolation. In fact, the lightest Higgs boson
mass reads as \cite{higgs,higgs2,higgs3}
\begin{equation}
m^2_h\simeq M^2_Z+\frac{3 g^2_2 m^4_t}{8 \pi^2
M^2_W}\left[\ln\left(\frac{M^2_S}{m^2_t}\right) +\frac{{{
X}_t^{\prime}}^2}{M^2_S}\left(1-\frac{{{ X}_t^{\prime}}^2}{12 M^2_S}\right)\right]
\end{equation}
where the mean stop mass-squared
\begin{equation}
M^2_S=\frac{1}{2}(m_{\widetilde{t}_{1}}^2+m_{\widetilde{t}_{2}}^2)
\end{equation}
is independent of $A_t^{\prime}$ while the left-right mixing term is
${X}_t^{\prime}=A_t-(\mu+{ A}_t^{\prime}) \cot \beta.$ Notice that the MSSM result \cite{Carena:2002es}
is recovered by setting $A_{t}^{\prime} \rightarrow 0$. For a clearer view of the impact
of $A_t^{\prime}$ on the Higgs boson mass, one notes that the
upper bound shift of the lightest Higgs boson mass can be expressed as
\begin{equation}
\Delta m^2_h \propto \mu ~ A_t^{\prime} ~ \cot \beta
\end{equation}
in which we have neglected the higher order $\cot \beta$ terms
since their effects will be negligible for the chosen $\tan \beta$
value, and the exact expression can easily be derived from (6).
This shift may vary from a few MeVs  (for low values of $|{
A}^{\prime}_t|$ ) up to tens of GeVs depending on the input
parameters. This is an important aspect since it modifies the
upper bound of the Higgs boson mass, and in case a Higgs signal
below $130\ {\rm GeV}$ is not observed at the LHC, it provides an
explanation for higher values of $m_h$ already in the MSSM
(without resorting to NMSSM or U(1)$^{\prime}$ models which
generically yield higher values for $m_h$).


\vspace{0cm}
\begin{figure}[htbp]
 \includegraphics[width=4in,height=2in]{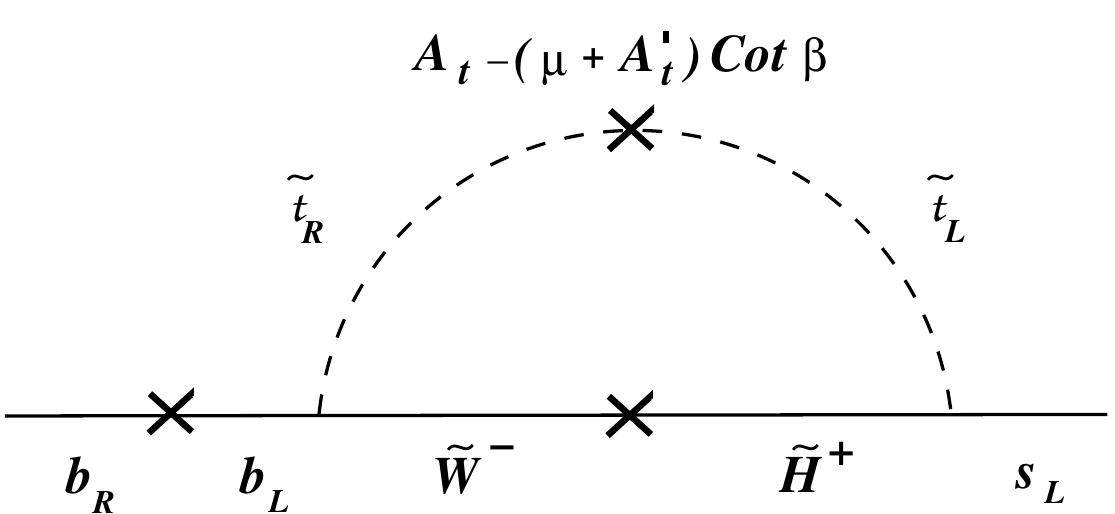}
   \vspace{0cm}
   \caption{The stop--chargino exchange contribution to $b
   \rightarrow s \gamma$ decay (photon can be coupled to any charged line).
   While the stop mixing is
   directly sensitive to $\mu+ A_t^{\prime}$ the chargino
   exchange involves mass of the charged Higgsino, the $\mu$
   parameter. This process thus involves both $\mu$ itself
   and $\mu+ A_t^{\prime}$ leading thus disentangling of
   $A_t^{\prime}$ from rest of the parameters. \label{fig1}}
\end{figure}

\begin{figure}[htbp]
 \vspace{-6.5cm}
\begin{center}
  \includegraphics[width=6in]{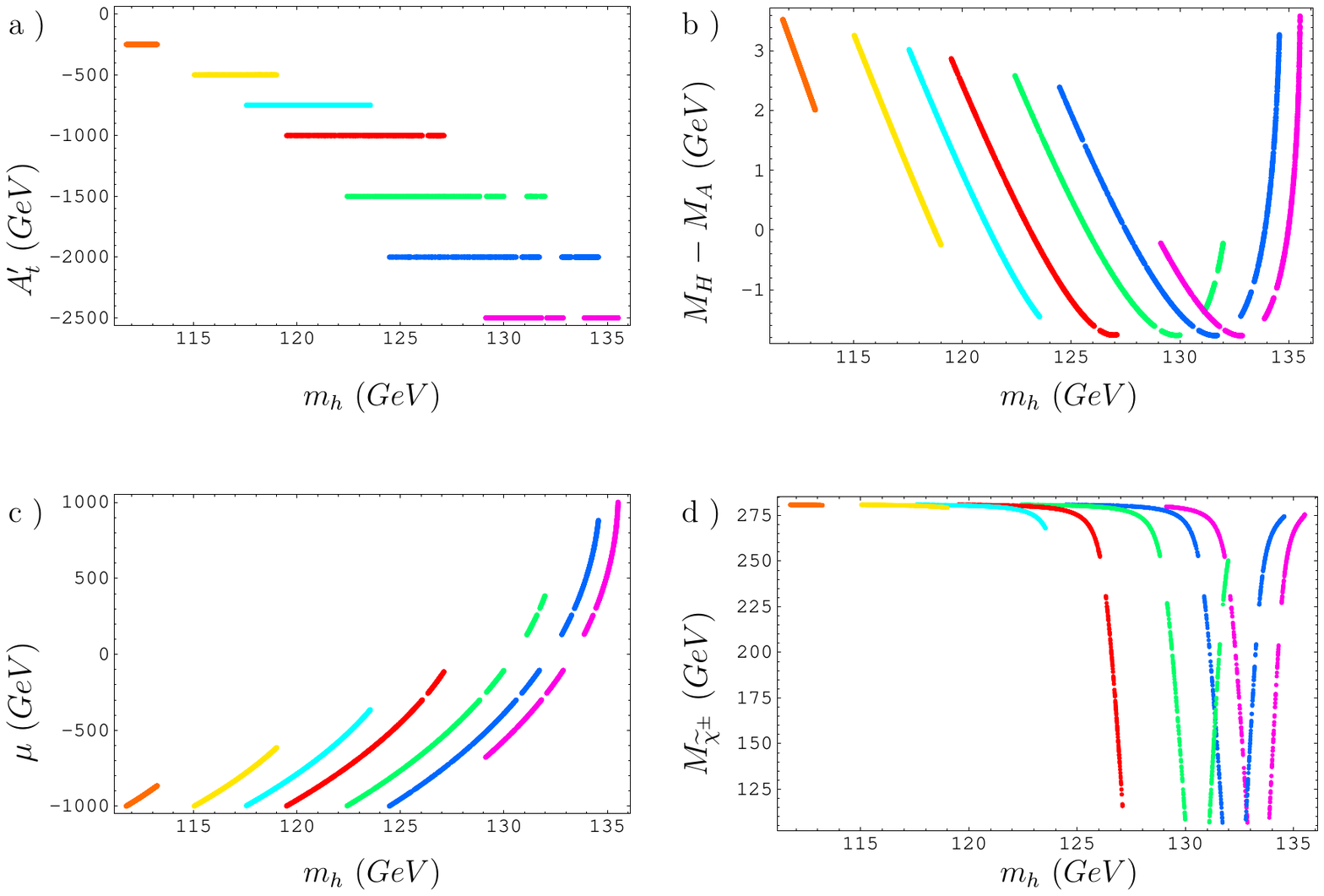}
   \vspace{-5cm}
   \caption{The lightest Higgs boson mass vs. certain model
   parameters after taking into account the $b \rightarrow s
   \gamma$ constraint. \label{fig2}}
   \end{center}
\end{figure}

The rare radiative decay $b \rightarrow s \gamma$ provides an
excellent arena for hunting the new physics effects since its
characteristic mass scale, the $b$ quark mass $m_b$, admits direct
application of perturbative QCD
\cite{bsgam1,bsgam2,bsgam3,bsgam4}. Moreover, experimental
precision has increased over the years at the level of essentially
confirming the SM result \cite{bsgam4,heavyflavor}. Therefore, the
branching ratio of this decay is expected to place rather
stringent limits on the sparticle contributions, and thus, provide
an almost unique way of determining the allowed ranges of
$A_t^{\prime}$. The reason behind this observation is that $b
\rightarrow s \gamma$ decay is sensitive to both $\mu$ (via
chargino exchange) and $\mu + A_t^{\prime}$ (via the stop
exchange) as illustrated in Fig. 1. Therefore, one has both $\mu$
and $\mu + A_t^{\prime}$ at hand simultaneously and thus it
becomes possible to disentangle $A_t^{\prime}$ effects from rest
of the soft masses. In fact, from the form of the chargino mass
matrix
\begin{eqnarray}
m_{\chi} =  \left(\begin{array}{c c}
M_2 & M_W \sqrt2 \sin\beta\\
M_W \sqrt2 \cos\beta & \mu
\end{array}\right)
\end{eqnarray}
one observes that wino and higgsino components mix due to
electroweak breaking (denoted by a cross on the horizontal line
inside the loop), and higgsino mass $\mu$ enters the branching
ratio in isolation. Unlike chargino sector, as suggested by Fig.
1,  the stop left-right mixing (denoted by a cross on the dashed
arc in the loop) depends explicitly on $\mu + A_t^{\prime}$ as
seen also from (\ref{stopmass}). The simultaneous $\mu$ and $\mu +
A_t^{\prime}$ dependencies of $b \rightarrow s \gamma$ decay, as
depicted in Fig. 1, results thus in a distinction between $\mu$
and $\mu + A_t^{\prime}$, which would not be possible by an
analysis of the Higgs sector alone.

Depicted in Fig.2 is the impact of the different $A_t^{\prime}$ values on certain parameters
respecting $b \rightarrow s \gamma$ restrictions.
The numerical results herein correspond to a specific choice of the parameters
\begin{eqnarray}
\label{param} M_1=140,\,M_2=280,\,M_3=1000,\,M_A=500,\\\nonumber
A_t=-1600,\,m_{t_L}=1000,\,m_{t_R}=200,\,
\end{eqnarray}
all in ${\rm GeV}$. We fix $\tan\beta = 5$ which is theoretically in accessible
region \cite{Schael:2006cr} and do not consider higher $\tan\beta$ values since
large $\tan \beta$ values reduce $A_{t}^{\prime}$ effects in this regime as also
can be seen from the left-right mixing entry of (\ref{stopmass}).
These parameter values are chosen judiciously such that $m_h$ and $\tan \beta$ agree with
the LEP II lower bound of $m_h \geq 114\ {\rm GeV}$ and  $\tan \beta > 2$  when $A_t^{\prime}$ vanishes \cite{Schael:2006cr}. This choice will help in revealing the effects of $A_t^{\prime}$ in a
transparent way. We will see that typically large negative values
of $A_t^{\prime}$ leads to observable changes where how large it
should be depends, of course, on the characteristic scale of soft
mass parameters in (\ref{stopmass}).

Fig.2 (a) shows our color convention and how $m_h$ depends on
$A_t^{\prime}$. It is seen that $m_h$ just agrees with the LEP bound
when $A_t^{\prime}$ is small in magnitude. However, as it grows in
negative direction up to $- 2.5\ {\rm TeV}$ the Higgs boson mass gets
gradually shifted towards the $135\ {\rm GeV}$ borderline. This clearly
shows that a measurement of the Higgs boson mass can imply strikingly
different parameter space than one would expect naively from a restricted
set of soft-breaking terms given in (\ref{soft}). In addition, the horizontal
behavior of the curves in Fig.2 (a) is due to the allowed range of $\mu$
parameter by the $b \rightarrow s \gamma$ bound. That is, the $\mu$
parameter takes on different values for each selection of
$A_t^{\prime}$ determined via the $b \rightarrow s \gamma$ restriction.
This is also reflected in Fig.2 (c).

Shown in Fig. 2 (b) is the mass splitting between the CP--odd and
CP--even Higgs bosons vs. the lightest Higgs boson mass. With the
presence of non-holomorphic trilinear coupling $A_{t}^{\prime}$
the CP-odd and CP-even higgs bosons may not be degenerate as they
are in the MSSM. It is clear that, for each value of
$A_t^{\prime}$ a corresponding splitting $\sim 3.5$ GeV can exist.
For small values of  $A_t^{\prime}$ the $\mu$ parameter falls in a
rather narrow band, that is, bigger the $A_t^{\prime}$ (in negative
direction) larger the range of $\mu$ parameter. This increase in the mass
splitting can be measured at the ILC if not at the LHC.

Depicted in Fig. 2 (c) is the dependence of $m_h$ on $\mu$
parameter for different values of $A_{t}^{\prime}$. At low
$A_{t}^{\prime}$ the $\mu$ parameter is preferred to be $- 1\ {\rm
TeV}$ for $m_h$ to agree with the experiment. However, as
$A_{t}^{\prime}$ grows to large negative values the $\mu$
parameter takes on its mirror symmetric value; $\mu = 1$ TeV.
This large swing in the allowed range of $\mu$ stems solely from
the dependence of the stop masses in (\ref{stopmass}) on $\mu$ and
$A_{t}^{\prime}$ where $b \rightarrow s \gamma$ does not allow
their sum to exceed a certain threshold due to rather narrow
band of values left to new physics contributions
\cite{bsgam4,heavyflavor}.

Finally, shown in Fig. 2 (d) is the variation of $m_h$ with the
lighter chargino mass $m_{\chi^{\pm}}$ as $A_{t}^{\prime}$ varies.
One notices how their relationship is modified at large negative
$A_t^{\prime}$ via especially the region at large $m_h$. Indeed,
as $A_t^{\prime}$ grows to large negative values the Higgs boson
mass is shifted towards $130\ {\rm GeV}$ border wherein change of
$m_{\chi^{\pm}}$ with $m_h$ is rather sharp. It is clear that both
these masses are measurable at the LHC, and their interdependence
can be guiding pivot if the model under concern is a minimal one based on
(\ref{soft}) or a more general one based on (\ref{soft}) and
(\ref{softp}) This of course requires a fit of the experimental data
to model parameters.

In principle, a full experiment on chargino and neutralino masses
must determine $M_2$, $M_1$, $\mu$ and $\tan\beta$ in a way
independent of what happens in the sfermion sector.
Experimentally, however, realization of this statement can be
quite non-trivial; in particular, one might need to determine
final states containing only neutralinos or only charginos or
neutralinos and charginos \cite{kane-measure}. An extraction of
$A_t^{\prime}$ then follows from constructing relations like the
ones illustrated in Fig. 2.

In conclusion, the soft breaking sector of the MSSM must in
general be extended to include also (\ref{softp}), and their
inclusion can have an important impact on various observables. In
particular, they influence radiative corrections to Higgs boson
masses, and their size can be examined within the LHC data by
forming a cross correlation among Higgs boson mass and other
observables. In particular, as illustrated in Fig. 2 (d), a
simultaneous knowledge of chargino and Higgs boson masses enables
one to search for $A_t^{\prime}$ effects after a fit to the model
parameters. The results advocated here could have important
implications for a global analysis of the LHC data.

\end{document}